# Unconventional and Powerful Ion Sources for Solid-State Ion Exchange, $Cu_2SO_4$ and $Cu_3PO_4$: Exemplified by Synthesis of Metastable β-$CuGaO_2$ from Stable β-$LiGaO_2$

Issei Suzuki[*,1], Kako Washizu[1], Daiki Motai[1], Masao Kita[2], and Takahisa Omata[1]

1. Institute of Multidisciplinary Research for Advanced Materials, Tohoku University, Sendai, Miyagi 980-8577, Japan
2. Department of Mechanical Engineering, National Institute of Technology, Toyama College, Toyama 939-8630, Japan

*Corresponding author: issei.suzuki@tohoku.ac.jp

**ABSTRACT:** This study introduces a new method for synthesizing $Cu^+$-containing metastable phases through ion exchange. Traditionally, CuCl has been used as a $Cu^+$ ion source for solid-state ion exchanges; however, its thermodynamic driving force is often insufficient for complete ion exchange with $Li^+$-containing precursors. First-principles calculations have identified $Cu_2SO_4$ and $Cu_3PO_4$ as more powerful alternatives, providing a higher driving force than CuCl. It has been experimentally demonstrated that these ion sources can open up new reaction pathways through experimental ion exchanges, such as from β-$LiGaO_2$ to β-$CuGaO_2$, which were previously unattainable. An important perspective provided by this study is that the potential of such basic compounds to act as powerful ion sources has been overlooked, and that they were identified through straightforward first-principles calculations. This work presents the initial strategic design of an ion exchange reaction by exploring suitable ion sources, thereby expanding the potential for synthesizing metastable materials.

Solid-state ion exchange is an efficient method for synthesizing metastable compounds. This technique entails topotactically substituting ions in a precursor that has the same or a similar crystal structure as the target materials while preserving the crystal framework, as illustrated by Reaction (1).[1-4]

AX (precursor) + BY (ion source) 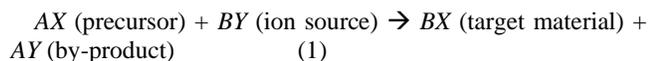 BX (target material) + AY (by-product)   (1)

Solid-state ion exchange generally occurs at relatively low temperatures (150–600 °C), preventing structural reconstruction into the most thermodynamically stable phase and enabling access to metastable phases not found on phase diagrams. In ion exchange, four chemical species are involved: the precursor and ion source (reactant system), as well as the target material and by-product (product system). The reaction is driven by the overall change in Gibbs free energy ($\Delta_r G = \Delta_r H - T\Delta_r S$).[4] A recent study has shown that the enthalpy change ($\Delta_r H$) of ion-exchange reactions, evaluated by first-principles calculations, can be used to screen whether the reaction will proceed, because the entropic gain due to ion mixing ($-T\Delta_r S$) is sufficiently small at the relatively low temperature.[5] A crucial yet previously underappreciated aspect of ion exchange is that even when the target material is fixed, $\Delta_r G$ of the overall reaction (i.e., whether the reaction will proceed or not) can be controlled by altering the combination of the reactant system. Conventional studies on ion exchanges, except for those involving $Ag^+$ ion exchange with $AgNO_3$, have predominantly utilized chloride salts such as CuCl or $CoCl_2$ as ion sources [3, 6-9] with no investigation into alternative ion sources. Consequently, if chloride salts failed to react with a particular precursor, further attempts to establish that reaction pathway were discontinued.[10] By strategically designing a new equilibrium field governed by four chemical species and exploring novel ion sources, researchers can significantly expand the accessible range of metastable materials.

In this study, we investigate ion exchange from $Li^+$-containing precursors to $Cu^+$-containing oxides as an illustrative example to explore new powerful ion sources through first-principles calculations. Additionally, we demonstrate that novel reaction pathways can be developed experimentally utilizing such ion sources.

In the synthesis of $Cu^+$-containing oxides through ion exchange, $Na^+$-containing precursors have primarily been utilized,[6, 7, 11, 12] while $Li^+$-containing precursors have been restricted to certain layered compounds.[8, 13] This limitation arises from the greater stability of $Li^+$-containing oxides compared to $Na^+$-containing oxides, coupled with the insufficient driving force for ion exchange provided by the CuCl ion source.[5] This contrast is evident in the ion exchange process for β-$CuGaO_2$: Reaction (2) involving a β-$NaGaO_2$ precursor, which exhibits a negative calculated $\Delta_r H$, results, in complete ion exchange to yield β-$CuGaO_2$. Conversely, Reaction (3) involving a β-$LiGaO_2$ precursor with a positive $\Delta_r H$ does not lead to successful ion exchange experimentally.[5]

β-$NaGaO_2$ + CuCl 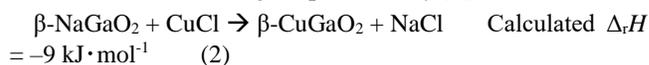 β-$CuGaO_2$ + NaCl    Calculated $\Delta_r H$ = –9 kJ·mol$^{-1}$   (2)

β-$LiGaO_2$ + CuCl 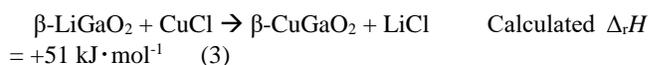 β-$CuGaO_2$ + LiCl    Calculated $\Delta_r H$ = +51 kJ·mol$^{-1}$   (3)

The ionic radius of $Na^+$ (1.00 Å for four-fold coordination) is significantly larger than those of $Cu^+$ and $Li^+$ (0.60 and 0.59 Å, respectively), leading to general challenges in ion exchanges from $Na^+$ to $Cu^+$, such as phase transition induced by coordination number changes and severe cracking due to volume shrinkage.[14, 15] Additionally, Na$^+$-containing precursors often suffer from Na deficiency owing to the high vapor pressure of Na,[16-18] resulting in severe cation deficiency in the obtained target materials (see detailed explanation in Section S1 in the supporting information).[14] These challenges can be addressed by using Li$^+$-containing precursors.

To identify powerful ion sources with ample driving force for ion exchange from Li$^+$ to Cu$^+$, we calculated the enthalpy difference between Cu$^+$-containing salts and their Li$^+$-containing counterparts. Using the identified Cu$^+$ ion source, an ion exchange pathway from β-LiGaO$_2$ to β-CuGaO$_2$ was demonstrated. This demonstration validates that investigating ion sources can unveil previously inaccessible ion exchange pathways.

The formation enthalpies of Cu$^+$-containing salts (CuCl, CuBr, CuI, Cu$_2$SO$_4$, Cu$_3$PO$_3$, CuCN, CuSCN, and CuH) and their Li$^+$-containing counterparts at 0 K were evaluated through first-principles calculations. Detailed calculation conditions and initial structures are provided in Section S2 in the supporting information. Ion exchange was experimentally conducted by mixing Li$^+$-containing precursor with the ion source in a Cu:Li = 1:1 ratio, followed by heating under vacuum (see details of synthesis of chemicals, and ion exchange process in Sections S3 and S4 in the supporting information). The reaction products were identified using X-ray diffraction (XRD, SmartLab, Rigaku, Japan), and their compositions were determined by dissolving the powder samples in a nitric acid solution in an autoclave, followed by inductively coupled plasma analysis (ICP, Optima 3300XL, Perkin Elmer, US).

Figure 1(a) presents the computed formation enthalpies of various Cu$^+$-containing salts and their corresponding Li$^+$-containing counterparts. Among the halide salts, CuCl exhibited the most negative $\Delta_rH$ for ion exchange, possibly elucidating why CuCl has traditionally been the preferred ion source in earlier ion exchange studies, given its easy availability. In contrast, utilizing Cu$_2$SO$_4$ and Cu$_3$PO$_4$ as ion sources yielded even more negative $\Delta_rH$ values. Specifically, the calculated driving forces of these salts were higher by 81 and 58 kJ·mol$^{-1}$, respectively, than that of CuCl. These robust driving forces are probably attributed to the metastable nature of Cu$_2$SO$_4$ and Cu$_3$PO$_4$.[19, 20]

To demonstrate that these ion sources are more powerful than CuCl, the synthesis of β-CuGaO$_2$ from β-LiGaO$_2$ described above should be a good case. Because the $\Delta_rH$ of β-LiGaO$_2$ and β-CuGaO$_2$ is +335 kJ·mol$^{-1}$, an ion source and by-product combination with a $\Delta_rH$ more negative than −335 kJ·mol$^{-1}$ is expected to drive the ion exchange (Figure 1(b)). The overall $\Delta_rH$ value for ion exchange using either Cu$_2$SO$_4$ or Cu$_3$PO$_4$ as ion sources (Reaction (4,5)) is −30.2 or −6.8 kJ·mol$^{-1}$, respectively, indicating potential reaction progression. However, this expectation is solely based on the thermodynamic perspective at 0 K. To complete ion exchange within a reasonable timeframe at experimental temperature, sufficiently high inter-diffusion coefficients of Cu$^+$ and Li$^+$ in these ion sources are necessary.[21] This was investigated using the following ion exchange experiment.

β-LiGaO$_2$ + $\frac{1}{2}$Cu$_2$SO$_4$ → β-CuGaO$_2$ + $\frac{1}{2}$Li$_2$SO$_4$     Calculated $\Delta_rH$ = −30 kJ·mol$^{-1}$  (4)

β-LiGaO$_2$ + $\frac{1}{3}$Cu$_3$PO$_4$ → β-CuGaO$_2$ + $\frac{1}{3}$Li$_3$PO$_4$     Calculated $\Delta_rH$ = −7 kJ·mol$^{-1}$  (5)

Figure 2(b,c) shows the XRD profiles of β-LiGaO$_2$ after heating with either CuCl or Cu$_2$SO$_4$. When CuCl was used as the ion source, β-LiGaO$_2$ remained unchanged, indicating no ion exchange, in line with a previous report.[5] Conversely, using Cu$_2$SO$_4$ as an ion source resulted in the formation of β-CuGaO$_2$ and the by-product Li$_2$SO$_4$, indicating successful ion exchange. The by-product was eliminated by water washing. The Cu$_2$O impurity likely originated from the partial decomposition of the Cu$_2$SO$_4$ ion source. During Cu$^+$ ion exchange, trace impurities such as Cu$_2$O and metallic Cu are commonly produced. These impurities can typically be eliminated by washing with aqueous ammonia, facilitating the isolation of the single-phase target material.[8] However, this method is unsuitable for β-CuGaO$_2$ due to its solubility in aqueous ammonia owing to the amphoteric nature of Ga. The chemical composition of the water-washed sample, as determined through the ICP analysis, was Li:Cu:Ga:S = 0.014:1:1.23:0.026. Although the Cu content could not be accurately determined owing to the presence of Cu$_2$O impurities, the low Li content suggests an almost complete replacement of Li$^+$ with Cu$^+$. Furthermore, the lattice constants of the resulting β-CuGaO$_2$ ($a_0$ = 5.472 Å, $b_0$ = 6.609 Å,

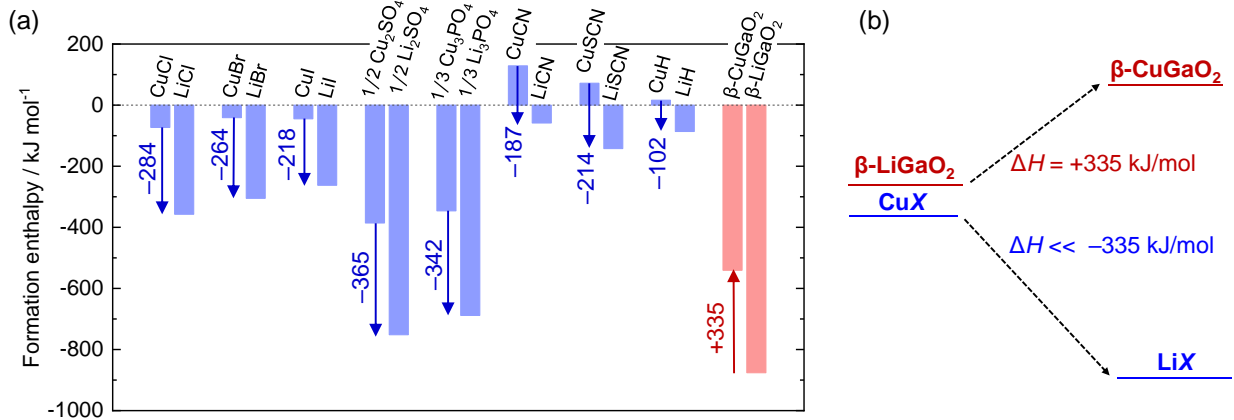

Figure 1. (a) Formation enthalpies of Cu$^+$-containing compounds and their Li$^+$-containing counterparts. (b) Schematic energy diagram of ion exchange between β-LiGaO$_2$ and Cu$^+$-containing ion source.

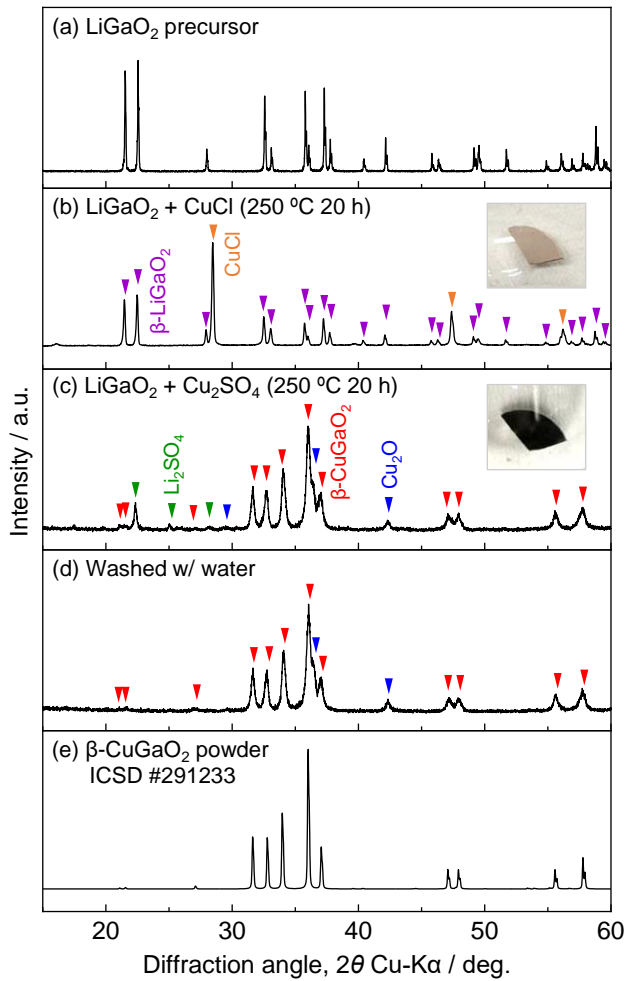

Figure 2. XRD profiles of the samples after the heating process: (b) β-LiGaO$_2$ and CuCl without washing, β-LiGaO$_2$ and Cu$_2$SO$_4$ (c) before and (d) after washing., along with (a) the profile of the β-LiGaO$_2$ precursor and (e) the simulated powder pattern of β-CuGaO$_2$ (ICSD 291233[22]). The inset show photographs of the samples.

and $c_0$ = 5.261 Å) closely matched the reported values ($a_0$ = 5.460 Å, $b_0$ = 6.610 Å, and $c_0$ = 5.274 Å),[22] further confirming nearly complete ion exchange. These results demonstrate that using Cu$_2$SO$_4$ as the ion source allows access to ion exchange that was unattainable with CuCl.

In the case of Cu$_3$PO$_4$ as an ion source (Reaction (5)), in contrast, β-LiGaO$_2$ remained unchanged after heating at 250 °C (See Section S5 in the supporting information). Additionally, the reverse reaction (i.e., the reaction between Li$_3$PO$_4$ and β-CuGaO$_2$) also did not proceed at 250 °C (See Section S6 in the supporting information). These results indicate that $\Delta_r H$ for this reaction is almost zero at the actual experimental temperature, considering the facts that uncertainty in determining the enthalpy of metal oxides by first-principles calculations has a standard deviation of 24 meV·atom$^{-1}$[23] (equivalent to 2.3 kJ·mol$^{-1}$ in this case) and that calculated $\Delta_r H$ is based on 0 K without considering temperature effects. Nevertheless, the complete ion exchange of Reaction (6) was experimentally achieved (Figure 3(a,b)). The direction of this reaction provides the direct evidence that Cu$_3$PO$_4$ functions as a more powerful ion source than CuCl.

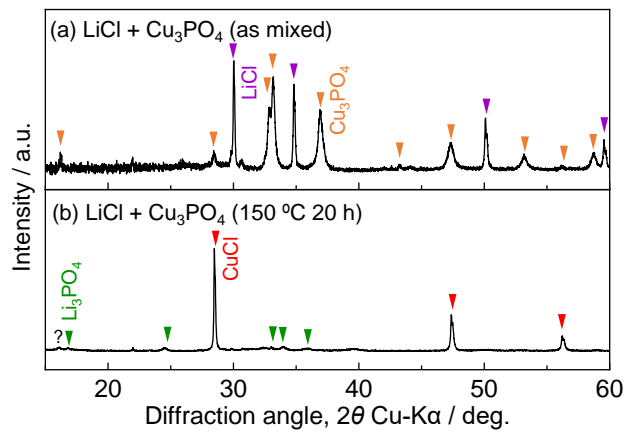

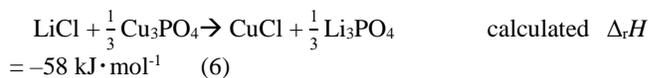

$\text{LiCl} + \frac{1}{3}\text{Cu}_3\text{PO}_4 \rightarrow \text{CuCl} + \frac{1}{3}\text{Li}_3\text{PO}_4$  calculated $\Delta_r H$ = –58 kJ·mol$^{-1}$   (6)

Figure 3. XRD profiles of LiCl and Cu$_3$PO$_4$ (a) before and (b) after heating.

Cu$_2$SO$_4$ is prone to instability in the air,[24] while Cu$_3$PO$_4$ offers enhanced atmospheric stability. Therefore, Cu$_3$PO$_4$ would be utilized as an easy-to-handle and more powerful Cu$^+$ ion source than CuCl, particularly in ion exchange processes where a driving force as high as that needed for β-LiGaO$_2$ is not required.

In summary, Cu$_2$SO$_4$ and Cu$_3$PO$_4$ were identified in this study as ion sources with stronger driving forces for ion exchange with Li$^+$-containing precursors compared to the conventional ion source CuCl. The use of Cu$_2$SO$_4$ facilitates ion exchange from stable β-LiGaO$_2$ to metastable β-CuGaO$_2$, a reaction pathway previously considered unachievable. While Cu$_3$PO$_4$ did not provide sufficient driving force for ion exchange with β-LiGaO$_2$, it is user-friendly and more powerful ion source than CuCl. It is highly intriguing that such basic compounds as Cu$_2$SO$_4$ and Cu$_3$PO$_4$ have been overlooked as powerful tools for inducing topotactic reactions, and that they were identified through straightforward first-principles calculations. Ion exchange is not limited to monovalent ions in oxides; it is also applicable to multivalent ions and other material groups, such as chalcogenides and pnictides,[25] implying that many powerful ion sources may remain undiscovered. This study is expected to serve as a starting point for accelerating further exploration of unconventional ion-exchange pathways and expanding the possibilities for synthesizing new inorganic metastable materials.

**Author Contributions**

**Issei Suzuki:** Conceptualization (lead), Investigation (equal), Computation (lead), Visualization (supporting), Funding acquisition(lead), Supervising (equal), Writing – original draft (lead). **Kako Washizu:** Investigation (equal), Visualization (lead). **Daiki Motai:** Investigation (supporting). **Masao Kita:** Investigation (supporting). **Takahisa Omata:** Supervising (equal), Writing – review and editing (lead).

**Notes**
The authors declare no competing financial interest.

## ACKNOWLEDGMENT

This work was partly supported by a Grant-in-Aid for Challenging Research (Exploratory) (Grant Nos. 22K18897 and 24K21680), the Research Program of "Five-star Alliance" in "NJRC Mater. & Dev.", and the TAGEN project from Institute of Multidisciplinary Research for Advanced Materials (IMRAM), Tohoku University.

Supporting information for

# Unconventional and Powerful Ion Sources for Solid-State Ion Exchange, $Cu_2SO_4$ and $Cu_3PO_4$: Exemplified by Synthesis of Metastable β-$CuGaO_2$ from Stable β-$LiGaO_2$

Issei Suzuki[*,1], Kako Washizu[1], Daiki Motai[1], Masao Kita[2], and Takahisa Omata[1]

1. Institute of Multidisciplinary Research for Advanced Materials, Tohoku University, Sendai, Miyagi 980-8577, Japan

2. Department of Mechanical Engineering, National Institute of Technology, Toyama College, Toyama 939-8630, Japan

**Section S1.** **Challenges associated with ion exchanges involving Na$^+$-containing precursors**

The ionic radius of Na$^+$ (1.00 Å for four-fold coordination) is notably larger than that of Cu$^+$ (0.59 Å). This difference poses challenges in ion-exchange reactions that aim to maintain the crystal framework of the Na$^+$-containing precursor. To address these challenges, it is advisable to utilize the Li$^+$-containing precursor since the ionic radius of Li$^+$ (0.60 Å) closely matches that of Cu$^+$.

(i) Na$^+$ tends to exhibit six-fold coordination rather than four-fold coordination in oxides due to its large ionic radius, while Cu$^+$ generally prefers two- or four-fold coordination. This distinction is evident in $LiMn_2O_4$ and $LiTi_2O_4$, which feature spinel-type structures with Li$^+$ in four-fold coordination, whereas the Na$^+$-containing counterparts ($NaMn_2O_4$ and $NaTi_2O_4$) adopt $CaFe_2O_4$-type structures with Na$^+$ in six-fold coordination.[1] Numerous instances in inorganic chemistry demonstrate this phenomenon, such as $LiVO_3$, where some Li$^+$ ions are four-fold coordinated, while all Na$^+$ ions in $NaVO_3$ are six-fold coordinated. Substituting six-fold coordinated Na$^+$ with Cu$_+$ would induce a change in the coordination environment and potentially lead to a phase transition, impacting the synthesis of the desired material (unless such a transition is intentionally sought).

(ii) Even when maintaining the coordination environment during the ion exchange from Na$^+$ to Cu$^+$, the disparity in ionic radii causes significant volume contraction. Reports indicate that transitioning from β-$NaGaO_2$ to β-$CuGaO_2$, whether in thin-film or single-crystalline forms, results in notable sample cracking. The use of a Li$^+$-containing precursor, with a lattice size typically closer to that of Cu$^+$-containing target materials, is anticipated to mitigate cracking. For example, Table S1 summarizes the lattice sizes of β-$CuGaO_2$, β-$NaGaO_2$, and β-$LiGaO_2$.

(iii) In addition to challenges related to differences in ionic radii, Na$^+$-containing oxides also encounter technical obstacles in achieving a stoichiometric composition without Na deficiency, primarily because of the inherently high vapor pressure of Na (e.g., 5×10$^5$ Pa at 800 °C[2]).[3-5] Since the total amount of cations before and after ion exchange remain the same, a precursor with a stoichiometric composition is essential to prevent cation deficiency after ion exchange. In contrast, the significantly lower vapor pressure of Li (2×10$^{-1}$ Pa at 800 °C[2]) facilitates the attainment of a stoichiometric composition in Li+-containing precursors.

Table S1. Lattice parameters of β-$CuGaO_2$[6], β-$NaGaO_2$[7] and β-$LiGaO_2$[8], along with the differences.

|  | β-$CuGaO_2$ | β-$NaGaO_2$ | β-$LiGaO_2$ |
| --- | --- | --- | --- |
| $a_0$ | 5.4600 | 5.498 (-0.70%) | 5.402 (-1.06%) |
| $b_0$ | 6.6101 | 7.206 (+9.01%) | 6.372 (-3.60%) |
| $c_0$ | 5.2742 | 5.298 (+0.45%) | 5.007 (-5.07%) |

**Section S2. Calculation conditions**

Calculations were performed on β-$M$GaO$_2$ ($M$ = Li$^+$, Cu$^+$), $M$Cl, $M$Br, $M$I, $M_2$SO$_4$, $M_3$PO$_3$, $M$CN, $M$SCN, and $M$H. The total enthalpies of these compounds were determined using first-principles calculations with the open-source Quantum Espresso software (version 5.2),[9, 10] with the Winmostar V11 GUI (X-Ability Co. Ltd., Japan). Formation enthalpies were obtained by calculating the total enthalpies of elemental substances (Cu, Ga, O$_2$, Cl$_2$, Br$_2$, I$_2$, S, P, C, N$_2$, and H$_2$). Projector-augmented wave-type pseudopotentials were utilized, generated with the 'atomic' code developed by Dal Corso (version 5.0.2) using scalar relativistic computations. The energy cutoffs for plane waves and charge density were set at 100 Ry (1.36 keV) and 900 Ry (12.3 keV), respectively. The self-consistent field (SCF) convergence threshold was set to 1 × 10$^{-7}$ Ry (1.36×10$^{-6}$ eV). Convergence thresholds for total energy, ionic minimization, and pressure in variable cell relaxation were 2×10$^{-5}$ Ry (2.7×10$^{-4}$ eV), 3×10-4 Ry·Bohr$^{-1}$ (1.2×10$^{-5}$ eV·Å$^{-1}$), and 12.5 MPa, respectively. The initial structures and geometry optimization parameters are provided in Table S2.

Table S2. Initial crystal structures used for geometric optimization in the DFT calculations obtained from either the ICSD database or the Materials Project (MP)[11]. The space group, $k$-point mesh, and calculated total enthalpies are also summarized.

| Composition | Initial structure | Space group | $k$-point | Total enthalpy / kJ·mol$^{-1}$ |
|---|---|---|---|---|
| β-LiGaO$_2$ | ICSD 18152[8] | Pna2$_1$ (33) | 5×4×5 | –489100.4552 |
| β-CuGaO$_2$ | ICSD 291233[6] | Pna2$_1$ (33) | 5×4×5 | –745406.1894 |
| LiCl | ICSD 27981 [12] | Fm-3m (225) | 4×4×4 | –122231.1863 |
| CuCl | ICSD 78270 [13] | F-43m (216) | 7×7×7 | –378588.0957 |
| LiBr | ICSD 52236 [14] | Fm-3m (225) | 4×4×4 | –283590.2986 |
| CuBr | ICSD 78274 [13] | F-43m (216) | 4×4×4 | –539967.6342 |
| LiI | ICSD 414244 [15] | Fm-3m (225) | 3×3×3 | –509373.7345 |
| CuI | ICSD 30363 [16] | P3m1 (156) | 3×3×1 | –765797.4729 |
| Li$_2$SO$_4$ | ICSD 2512 [17] | P12$_1$/c1 (14) | 2×3×2 | –341222.1799 |
| Li$_2$SO$_4$ | ICSD 153806 [18] | Cmcm (63) | 3×3×2 | –341195.9445 |
| Cu$_2$SO$_4$ | ICSD 40452 [19] | FdddZ (70) | 2×2×2 | –853773.1019 |
| Li$_3$PO$_4$ | ICSD 10257 [20] | Pmn2$_1$ (31) | 2×3×3 | –347395.7212 |
| Li$_3$PO$_4$ | ICSD 77095 [21] | Pnma (62) | 1×2×3 | –347393.8068 |
| Cu$_3$PO$_4$ | ICSD 427086 [22] | P-3 (147) | 5×4×4 | –1116292.631 |
| LiCN | ICSD 77321 [23] | Pnma (62) | 2×4×2 | –79922.53618 |
| CuCN | MP 35308 | R-3m (166) | 3×3×3 | –336376.5932 |
| LiSCN | ICSD 425061 [24] | Pnma (62) | 1×4×3 | –165329.6032 |
| CuSCN | ICSD 32578 [25] | P63mc (186) | 4×4×1 | –421756.9049 |
| LiH | ICSD 61749 [26] | Fm-3m (225) | 5×5×5 | –21179.32676 |
| CuH | ICSD 44859 [27] | P6$_3$mc | 9×9×6 | –277718.1794 |

## Section S3. Synthesis of chemicals

### *β-LiGaO₂*

β-LiGaO$_2$ powder was synthesized by Reaction (S1) from Li$_2$CO$_3$ (99%, Fujifilm Wako, Japan) and Ga$_2$O$_3$ (99.9%, Kojundo Chemical, Japan): [28]

$$\text{Li}_2\text{CO}_3 \text{ (s)} + \text{Ga}_2\text{O}_3 \text{ (s)} \rightarrow \text{LiGaO}_2 \text{ (s)} + \text{CO}_2 \text{ (g)} \quad\quad (S1)$$

Li$_2$CO$_3$ and Ga$_2$O$_3$ were weighed in a ratio of Li:Ga = 1.06:1 to consider Li evaporation, then mixed using a planetary ball mill. The mixture was formed into pellets through uniaxial pressing at 100 MPa and calcined at 600 °C for 48 h in the air. The resulting sample underwent washing with ultrapure water and ethanol to eliminate any unreacted Li$_2$CO$_3$. The powder was subsequently re-mixed with the planetary ball mill, re-pelletized, and sintered at 1100 °C for 28 h in air. Throughout the sintering process, the pellets were immersed in a mixture of Li$_2$CO$_3$ and Ga$_2$O$_3$ powder to reduce Li evaporation.

### *Cu₂SO₄*

Cu$_2$SO$_4$ was synthesized by Reaction (S2) using fine-particle Cu$_2$O (FRC-05B, Furukawa Chemicals, Japan) and (CH$_3$)$_2$SO$_4$ (dimethyl sulfate, Fujifilm Wako Pure Chemicals, Japan).[19] The Cu$_2$O used consisted of extremely fine particles, consistent with previous reports that fine-particle Cu$_2$O is desirable for this synthesis method. [19]

$$\text{Cu}_2\text{O (s)} + (\text{CH}_3)_2\text{SO}_4 \text{ (}\ell\text{)} \rightarrow \text{Cu}_2\text{SO}_4 \text{ (s)} + (\text{CH}_3)_2\text{O (g)} \quad\quad (S2)$$

A 10 mL of (CH$_3$)$_2$SO$_4$ and 0.1 g of Cu$_2$O were combined in a round-bottom flask and stirred at 250 rpm under an Ar atmosphere. The flask was immersed in a 160 °C oil bath and heated for 10 minutes while stirring. Subsequently, the flask was taken out from the oil bath and cooled rapidly in room-temperature water. The supernatant of liquid was decanted, and the precipitate was dried under a vacuum at room temperature.

The acquired Cu$_2$SO$_4$ was verified as a single phase in the XRD profile (Figure S1). This Cu$_2$SO$_4$ can be preserved in vacuum desiccator or under an inert atmosphere at room temperature for a minimum of several weeks without deterioration.

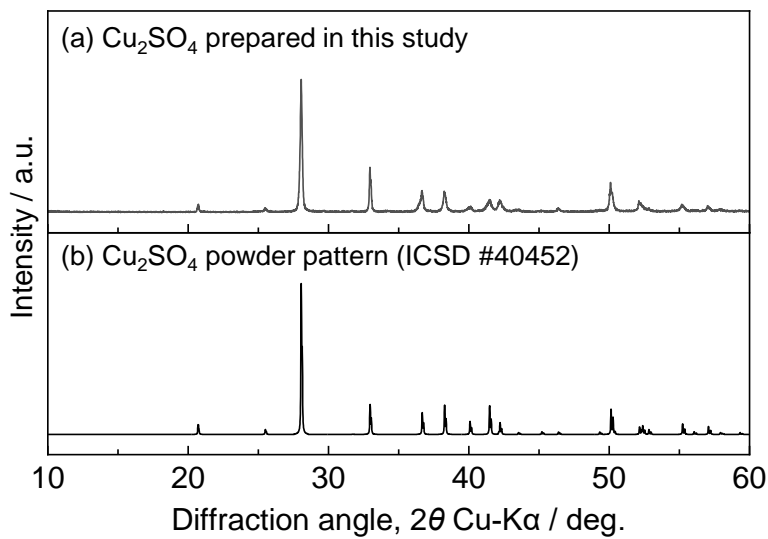

Figure S1. XRD profiles of the obtained $Cu_2SO_4$ compared with the simulated powder pattern of $Cu_2SO_4$ (ICSD 40452 [19]).

### $Cu_3PO_4$

$Cu_3PO_4$ was synthesized via a two-step Reactions (S3, S4).[22]

$$2(NH_4)_2HPO_4 \text{ (s)} + 3CuO \text{ (s)} \rightarrow Cu_3(PO_4)_2 \text{ (s)} + 4NH_3 \text{ (g)} + 3H_2O \text{ (g)} \quad \text{(S3)}$$

$$Cu_3(PO_4)_2 \text{ (s)} + 3Cu \text{ (s)} \rightarrow 2Cu_3PO_4 \quad \text{(S4)}$$

($(NH_4)_2HPO_4$ (99.0+%, Fujifilm Wako Pure Chemical) and CuO (99.9%, Fujifilm Wako Pure Chemical) were weighed in a 2:3 molar ratio and mixed with air using a mortar and pestle for 30 minutes. The mixture was then uniaxially pressed at 100 MPa to form Φ9 mm pellets, which were sintered in air at 1000 °C for 40 h to obtain single-phase $Cu_3(PO_4)_2$ (Figure S2(a)). The resulting $Cu_3(PO_4)_2$ was then mixed with metallic Cu powder (99%, ~75 μm, Fujifilm Wako Pure Chemicals) in a 1:3 molar ratio using a mortar and pestle in a glove box filled with a $N_2$ atmosphere for about 30 min. This mixture was then pressed into Φ 9 mm pellets under 100 MPa and vacuum-sealed in a quartz tube. The tube was heated at 850 °C for 72 h and then rapidly quenched by placing it in water. This process yielded nearly single-phase $Cu_3PO_4$ (Figure S2(c)). This $Cu_3PO_4$ can be preserved in vacuum desiccator or under an inert atmosphere at room temperature for a minimum of several months without deterioration.

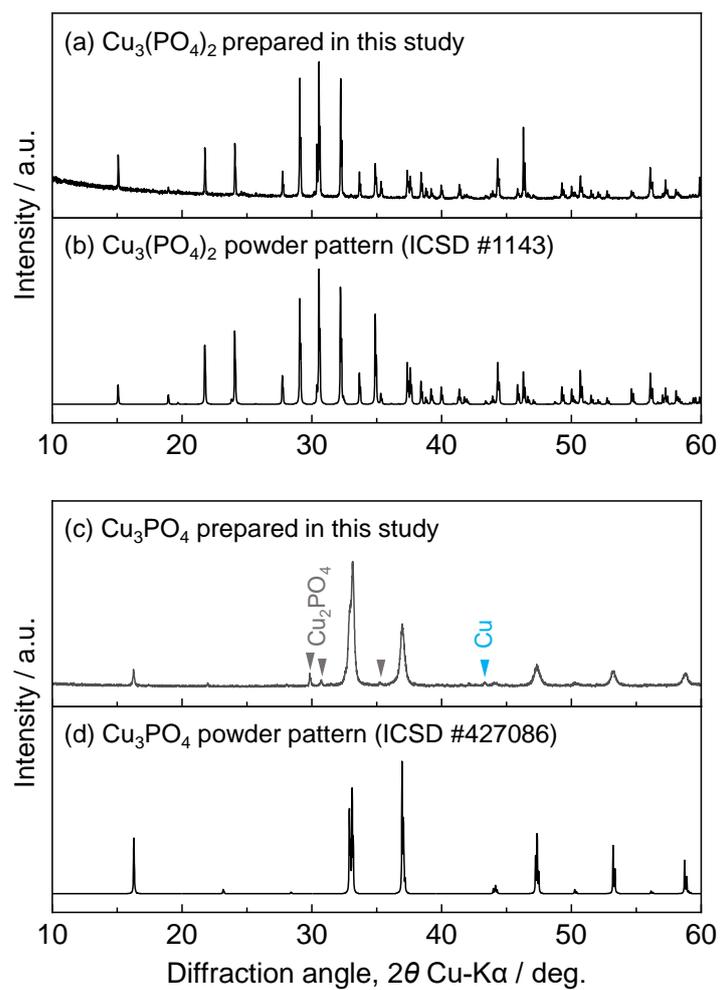

Figure S2. XRD profiles of (a) $Cu_3(PO_4)_2$ and (c) $Cu_3PO_4$ synthesized in this study, along with the simulated powder patterns of $Cu_3(PO_4)_2$ (ICSD #1143 [29]) and $Cu_3PO_4$ (ICSD #427086 [22]).

## Section S4. Ion-exchange processes

### *β-LiGaO$_2$ and Cu$_2$SO$_4$ or Cu$_3$PO$_4$*

β-LiGaO$_2$ powder and Cu$_2$SO$_4$ or Cu$_3$PO$_4$ powder were mixed in a ratio of Li:Cu = 1:1 in a glove box filled with N$_2$ gas using a mortar and pestle for 15 min. Subsequently, the homogenized mixture was pressed into a Φ9 mm pellet at 100 MPa using a uniaxial press. The pellet was positioned at the base of a Pyrex test tube and subjected to heating at either 150 or 250 °C for 20 h in an electric furnace, while the opposite end of the test tube was connected to a rotary pump. Following the reaction, the sample was rinsed with ultrapure water, and the resulting precipitate was harvested via centrifugation. This washing procedure was iterated thrice, culminating in a final rinse with ethanol. The precipitate, post-final centrifugation, was subsequently dried in a vacuum desiccator.

### *LiCl and Cu$_3$PO$_4$*

LiCl powder (99.9%, Fujifilm Wako) was blended with Cu$_3$PO$_4$ in a ratio of Li:Cu = 1:1 and underwent an ion exchange process similar to the method employed for β-LiGaO$_2$ above. Heating temperature and duration were 150 °C and 20 h, respectively.

## Section S5. Reaction of β-LiGaO$_2$ and Cu$_3$PO$_4$

When Cu$_3$PO$_4$ was utilized as the ion source, no reaction occurred upon heating to 150 °C, except for the formation of Cu and Cu$_2$PO$_4$ due to partial disproportionation of the ion source (Figure S3(a)). Increasing the reaction temperature to 250 °C led to complete disproportionation of Cu$_3$PO$_4$, while β-LiGaO$_2$ remained unaffected.

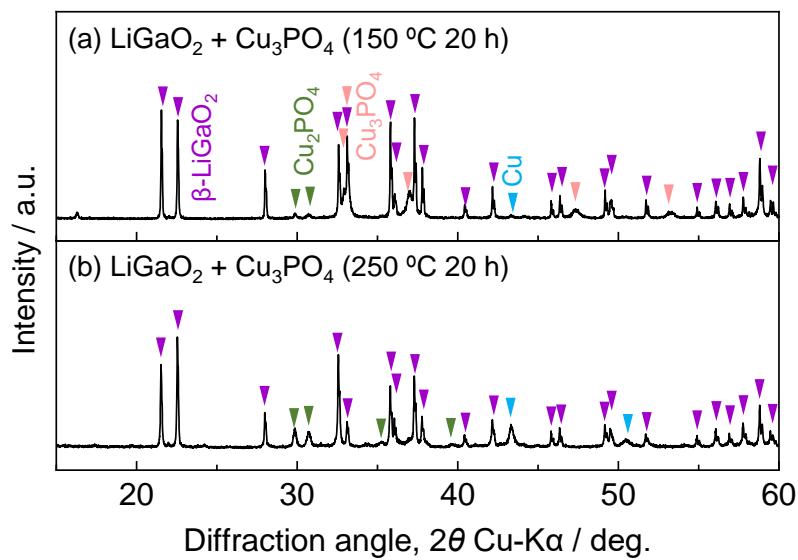

Figure S3. XRD profiles of the samples after the heating process: β-LiGaO$_2$ and Cu$_3$PO$_4$ at (a) 250 °C and (b) 200 °C, and (c) β-CuGaO$_2$ and Li$_3$PO$_4$ at 250 °C.

## Section S6. Reaction of β-CuGaO$_2$ and Li$_3$PO$_4$

As the reverse reaction of Reaction (5) in the main text, Reaction (S5) was conducted. β-CuGaO$_2$ was synthesized through ion exchange between β-NaGaO$_2$ and CuCl, as previously documented.[6] β-CuGaO$_2$ and commercially available Li$_3$PO$_4$ (95%, Fujifilm Wako, Japan) were weighed in a ratio of Cu:Li = 1:1, mixed, and then compacted into a pellet. The pellet was heated under vacuum at 250 or 350 °C for 20 h using the same setup as detailed for the ion exchange involving β-LiGaO$_2$ in Section S4.

$$\beta\text{-CuGaO}_2 + \tfrac{1}{3}\text{Li}_3\text{PO}_4 \rightarrow \beta\text{-LiGaO}_2 + \tfrac{1}{3}\text{Cu}_3\text{PO}_4 \quad \Delta_r H = +6.8 \text{ kJ·mol}^{-1} \quad (S5)$$

As shown in Figure S4, while heating β-CuGaO$_2$ and Li$_3$PO$_4$ at 250 °C did not result in any change to β-CuGaO$_2$, heating at 350 °C yields XRD peak shifts. Since the Vegard's law holds for the lattice parameters of the solid solution of β-CuGaO$_2$ and β-LiGaO$_2$,[30] the composition evaluated of the sample from lattice constants was $x$ =0.12 (Li$_{0.12}$Cu$_{0.88}$GaO$_2$), as shown in Figure S5. The formation of this solid solution indicates that the increase in entropy gain (–$T\Delta S$) with increasing temperature acted as the primary driving force for ion exchange, supporting the assumption that the $\Delta_r H$ of Reactions (5) and (S5) are almost zero.

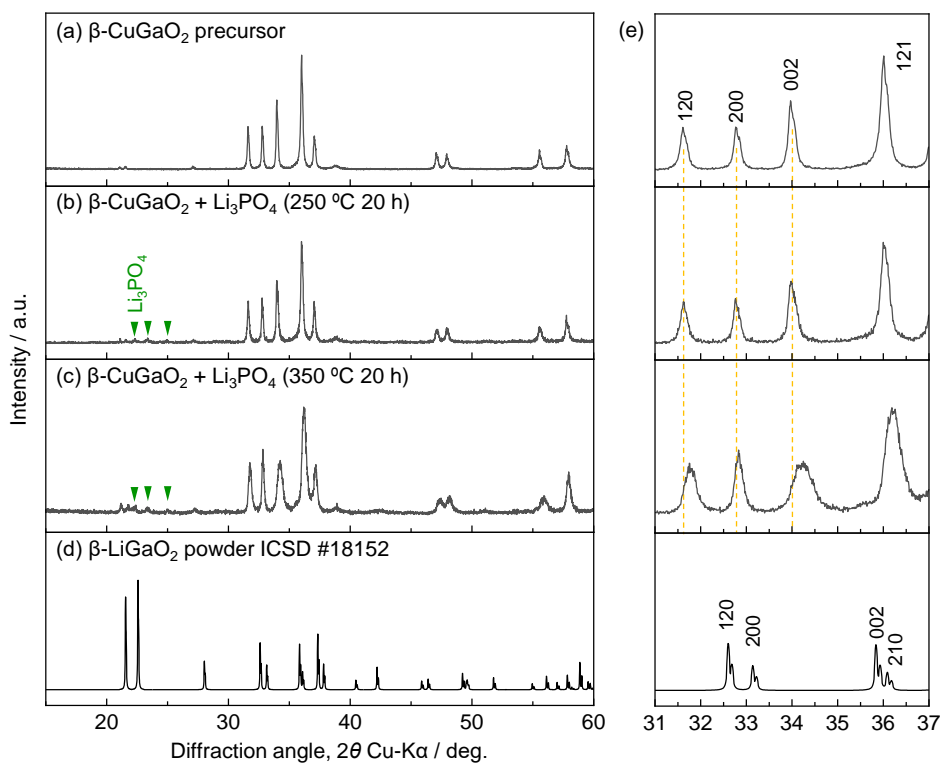

Figure S4. XRD profiles of the sample after the heating process of β-CuGaO$_2$ with Li$_3$PO$_4$ (b) at 250 °C and (c) at 350 °C for 20 h, along with the patterns of (a) β-CuGaO$_2$ precursor and (d) β-LiGaO$_2$ powder (ICSD #18152 [8]). (e) Enlarged profiles from 31 to 37°.

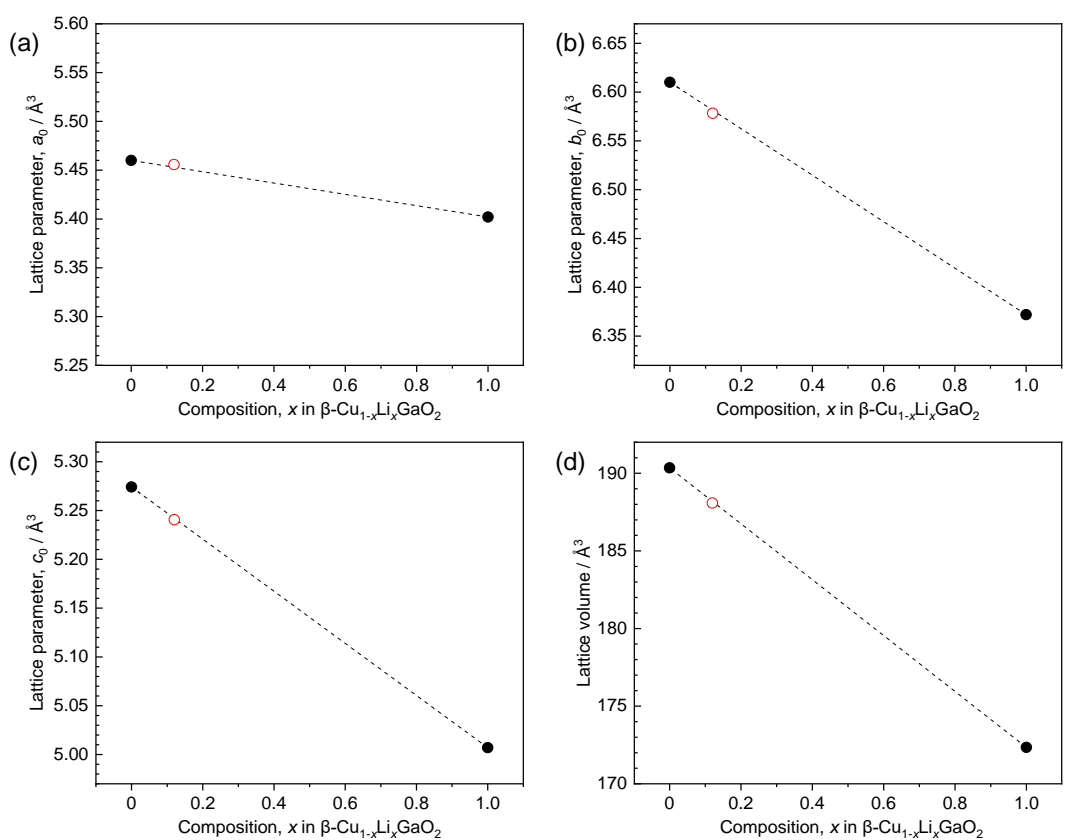

Figure S5. Compositional dependence of (a–c) lattice parameters and (d) lattice volume. The terminals are literature values of β-CuGaO$_2$[6] and β-LiGaO$_2$[8]. The composition of the sample after heating at 350 °C (red open circle figures) corresponds to $x$ = 0.12 (Li$_{0.12}$Cu$_{0.88}$GaO$_2$).

# References for supporting information